\documentclass{aastex}
\usepackage{emulateapj5}
\lefthead{TSUJIMOTO \& SHIGEYAMA}
\righthead{History of $\omega$ Centauri}

\def\Msun{M_{\odot} }
\def\cm3{{\rm ~cm}^{-3}}
\def\ltsima{$\; \buildrel < \over \sim\;$}
\def\ltsim{\lower.5ex\hbox{\ltsima}}
\def\gtsima{$\; \buildrel > \over\sim \;$}
\def\gtsim{\lower.5ex\hbox{\gtsima}}
\def\ms{$M_{\odot}$ }
\def\msp{$M_{\odot}$}

\begin{document}
\title{Star Formation History of $\omega$ Centauri Imprinted in
Elemental Abundance Patterns}

\author{Takuji Tsujimoto$^{1}$ and Toshikazu Shigeyama$^{2}$}

\altaffiltext{1}{National Astronomical Observatory, 2-21-1 Ohsawa,
Mitaka-shi, Tokyo 181-8588, Japan; taku.tsujimoto@nao.ac.jp}

\altaffiltext{2}{Research Center for the Early Universe, Graduate
School of Science, University of Tokyo, 7-3-1 Hongo, Bunkyo-ku, Tokyo
113-0033, Japan; shigeyama@resceu.s.u-tokyo.ac.jp}

\begin{abstract}
The star formation history of the globular cluster $\omega$ Centauri
is investigated in the context of an inhomogeneous chemical evolution
model in which supernovae induce star formation. The proposed model
explains recent observations for $\omega$ Cen stars, and divides star
formation into three epochs. At the end of the first epoch, $\sim
70$\% of the gas was expelled by supernovae. AGB stars then supplied
$s$-process elements to the remaining gas during the first interval of
$\sim 300$ Myr. This explains the observed sudden increase in Ba/Fe
ratios in $\omega$ Cen stars at [Fe/H]$\sim -1.6$. Supernovae at the
end of the second epoch were unable to expel the gas. Eventually, Type
Ia supernovae initiated supernova-induced star formation, and
remaining gas was stripped when the cluster passed through the newly
formed disk of the Milky Way. The formation of $\omega$ Cen is also
discussed in the framework of globular cluster formation triggered by
cloud-cloud collisions. In this scenario, the relative velocity of
clouds in the collision determines the later chemical evolution in the
clusters. A head-on collision of proto-cluster clouds with a low
relative velocity would have converted less than 1\% of gas into stars
and promoted the subsequent chemical evolution by supernova-driven
star formation. This is consistent with present observed form of
$\omega$ Cen. In contrast, the other Galactic globular clusters are
expected to have formed from more intense head-on collisions, and the
resultant clouds would have been too thin for supernovae to accumulate
enough gas to form the next generation of stars. This explains the
absence of chemical evolution in these other globular clusters.
\end{abstract}

\keywords{Galaxy: halo --- globular clusters: general --- globular
clusters: individual ($\omega$ Centauri) --- stars: abundances ---
supernovae: general --- supernova remnants}

\section{INTRODUCTION}

The potential history and characteristics of the globular cluster (GC)
$\omega$ Centauri remains largely unresolved, although its most
conspicuous features can be seen in the elemental abundance
properties. Most prominently, there exists a large dispersion in
metallicity (--2\ltsim [Fe/H]\ltsim --0.5) among its member stars
\citep[e.g.,][]{Norris_95, Suntzeff_96} in contrast to other Galactic
GCs which exhibit essentially no dispersion. In addition, this GC
contains stars that display a significant enhancement of $s$-process
elements relative to iron (Fe) and $r$-process elements
\citep{Norris_95}.

The spread in metallicity (or [Fe/H]) is likely to be evidence of
self-enrichment \citep[e.g.,][]{Suntzeff_96}, and has been generally
attributed to the fact that $\omega$ Cen is the most massive and
brightest GC in the Galaxy \citep[e.g.,][]{Ikuta_00}. However,
\citet{Gnedin_02} showed that $\omega$ Cen does not have an
exceptionally deep gravitational potential compared to other Galactic
GCs. The uniqueness of $\omega$ Cen has led some authors to consider
that it is the surviving nucleus of a dwarf galaxy that has been
tidally stripped by the Milky Way
\citep[e.g.,][]{Freeman_93,Majewski_00,Hilker_00}. It should be noted
that there are no observed metal-poor ([Fe/H]\ltsim --2) stars in
$\omega$ Cen, which indicates that $\omega$ Cen formed from a
pre-enriched gas. This feature is indeed shared with the other
GCs. The formation of $\omega$ Cen should therefore be considered in
any general scheme of GC formation.

Regarding the latter feature, a study by \citet{Smith_00} identified
two distinct populations in terms of correlation between [Ba/Eu] and
[Fe/H] for 10 $\omega$ Cen stars ({\it filled circles} in Fig.~1). One
population has metallicity in the range of [Fe/H]$<-1.8$ and low
[Ba/Eu] ratios, corresponding closely to the pure $r$-process origin
ratio, and the other population has [Fe/H]$>-1.6$ with high [Ba/Eu]
ratios, ascribed to a large $s$-process nucleosynthesis
contribution. Another unique feature of the elemental abundance can be
observed in the abundance distribution function (ADF) of $\omega$ Cen
stars \citep{Norris_96, Suntzeff_96, van Leeuwen_00}. The ADF has a
conspicuous peak at [Fe/H]$\sim -1.4$ with a long, metal-rich tail
starting from [Fe/H]$\sim -1$ extending to [Fe/H]$\sim-0.5$. It is
expected that both the unusual ADF and the peculiar behavior of
$n$-capture elements could be useful in inferring the potentially
unique history of $\omega$ Cen.

\begin{figure*}[htb]
\begin{center}
\includegraphics[width=12cm,angle=0]{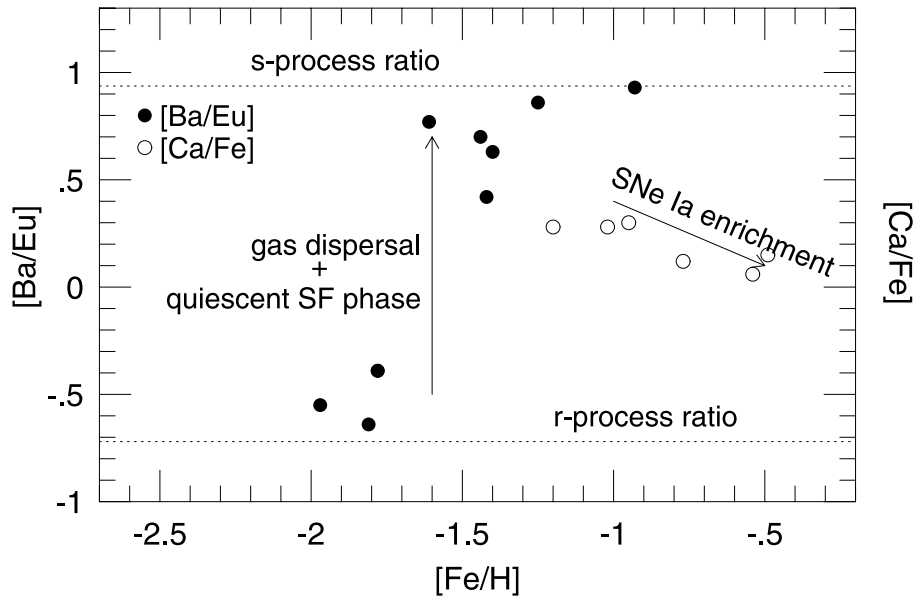}
\end{center}
\caption{Correlations of [Ba/Eu]\citep[{\it filled
circles};][] {Smith_00} and [Ca/Fe]\citep[{\it open
circles};][]{Pancino_02} with [Fe/H] for $\omega$ Cen stars. The
dashed lines show the [Ba/Eu] ratios theoretically anticipated from
the $r$-process origin and $s$-process nucleosynthesis. Some comments
are also attached.}
\end{figure*}

Recent observations have revealed that stars belonging to the
metal-rich tail exhibit signs of Type Ia supernovae (SNe Ia)
enrichment \citep{Pancino_02}. Figure 1 shows that the [Ca/Fe] ratios
({\it open circles}) decrease with increasing [Fe/H]. This gives an
overall sequence of chemical enrichment processes in $\omega$ Cen,
from initiation by Type II SNe ($r$-process), to an $s$-process, and
finally SNe Ia. The imprint of such a sequential enrichment history on
these stars strongly supports a self-enrichment scenario above among
several other possible explanations for the abundance anomalies of
$\omega$ Cen stars proposed so far \citep[see][]{Kraft_79,Smith_87}.

\citet[TS02 in the following]{Tsujimoto_02} have shown that the
stellar abundance pattern of $n$-capture elements could be used as a
powerful tool to infer how star formation proceeded in the Milky Way
dwarf spheroidal galaxies. In the same manner, the unique elemental
pattern seen in $\omega$ Cen stars must record the history of $\omega$
Cen. Combining all the above elemental abundance information, it is
possible to build a new picture for the history of $\omega$ Cen. The
scenario presented here is constructed in the context of an
inhomogeneous chemical evolution model in which stars are born from
the matter swept up by individual supernova remnants (SNRs)
\citep{Shigeyama_98}. In this model, each star inherits heavy elements
not only through the interstellar medium but also directly from a
massive star of the preceding generation. This model successfully
reproduces the chemical evolution of Galactic halo field stars
\citep{Tsujimoto_99} as well as that of the Milky Way dwarf spheroidal
stars (TS02), and is extended here to Galactic GCs including
$\omega$ Cen.

In the next section, the history of $\omega$ Cen is described based on
an analysis of the elemental abundance pattern of $\omega$ Cen
stars. The chemical evolution of $\omega$ Cen is modeled in section 3,
and a scheme for GC formation in the framework of the inhomogeneous
chemical evolution of the Galactic halo is proposed in section 4. The
paper is concluded in section 5.

\section{History of $\omega$ Centauri}

The star formation history of $\omega$ Cen is inferred from the
stellar elemental abundance pattern shown in Figure 1. First of all, a
sudden increase in the [Ba/Eu] ratios at [Fe/H]$\sim -1.6$ implies
that $\omega$ Cen has undergone two discrete episodes of star
formation. The observed [Ba/Eu] ratios of as large as +1 are
attributed to a contribution from asymptotic giant branch (AGB) stars
after gas removal at the end of the first episode of star
formation. If gas was not expelled after the first episode, the
$s$-process nucleosynthesis in AGB stars would increase the [Ba/Eu]
ratio to only $\sim +0.2$ at most according to the Salpeter initial
mass function (IMF). A steepened IMF may enhance the [Ba/Eu]
ratios. However, the observed [$\alpha$-elements/Fe] ratios for [Fe/H]
\ltsim --1 in $\omega$ Cen stars close to those for the solar
neighborhood stars \citep{Norris_95, Smith_00} require one similar to
the Salpeter IMF.

The star formation history inferred from the observed [Ba/Eu] ratios
is then as follows. 1. After completion of the first phase of star
formation, a large part of the remaining gas was expelled by SN II
explosions. 2. The gas was then contaminated by $s$-process elements
ejected from AGB stars during a quiescent stage lasting several
hundred Myr. This led to the enhancement of Ba relative to Eu in the
gas. 3. Star formation then recommenced after the gas, once heated by
SNe, cooled sufficiently and collapsed, with associated shock wave
formation. Stars formed from this shocked gas and stars subsequently
formed have very high Ba/Eu ratios. The second star formation phase
continued until the metallicity [Fe/H] reached $\sim -1$.

In this way, the second episode of star formation in $\omega$ Cen
should have proceeded without expelling all the interstellar gas. The
fate of the interstellar gas, i.e., whether it was expelled or
trapped, is determined by competition between the energy of SN
explosions and the gravitational binding energy of the gas. To
reproduce the observed features in $\omega$ Cen, the energy of SN
explosions should not appreciably exceed the gravitational binding
energy. A chemical evolution model that satisfies this condition is
shown in the next section. In the Milky Way dwarf spheroidal galaxies,
the absence of $s$-process signatures suggests that the energy of SN
explosions will have overcome the gravitational potential energy in
those cases, which prevents recommencement of star formation (TS02).

This then poses the question of how star formation proceeded after the
second star formation epoch. A clue lies in the extended metal-rich
tail in the observed ADF (Fig.~2a). \citet{Pancino_02} recently found
that stars belonging to the metal-rich tail exhibit decreasing [Ca/Fe]
ratios with increasing [Fe/H], ascribed as a sign of SNe Ia enrichment
({\it open circles} in Fig.~1). From the theoretical aspect, such SNe
Ia enrichment is indispensable to increasing the metallicity [Fe/H] of
stars to $\sim -0.5$ as observed. In fact, this metal rich end of
[Fe/H] cannot be reached by the Fe supply from only SNe II given the
small number of stars with [Fe/H] $>-1$ (see Fig.~2a). In the
SN-induced star formation scenario, SNe Ia do not only enrich gas with
Fe but also initiate star formation in the third episode.  

In the end, $\omega$ Cen has three episodes of star formation, each of
which is likely to be identified with each subpopulation found by
\citet{Pancino_00}. In addition, \citet{Ferraro_02} found that three
distinct populations have different proper-motion distributions: the
other two populations have bulk motions with respect to the dominant
metal-poor population (i.e., stars produced in the first episode).
During the star formation phase over a few Gyrs, the proto-cluster
cloud for $\omega$ Cen passed repeatedly through the forming Galactic
disk. At that time, the gas in the cloud was accelerated by the ram
pressure, leading to a bulk motion of the gas from which stars were
formed. In this framework, it is expected that the bulk motion of
third metal-rich population is larger than that of the second
population. The increasing density of the disk exerted a greater ram
pressure to the proto-cluster cloud at a later passage leading to a
larger bulk motion. It is indeed consistent with the observed
finding. The ram pressure associated with the passages of the disk was
not strong enough to strip the gas off, that is, the ram pressue could
not overcome the gravitational attraction of the cloud, as discussed
in \S 4.3. Thus there might be no need to invoke an accretion of
metal-rich stellar system onto $\omega$ Cen as suggested by
\citet{Ferraro2_02}.

\begin{figure*}[htb]
\begin{center}
\includegraphics[width=10cm,angle=0]{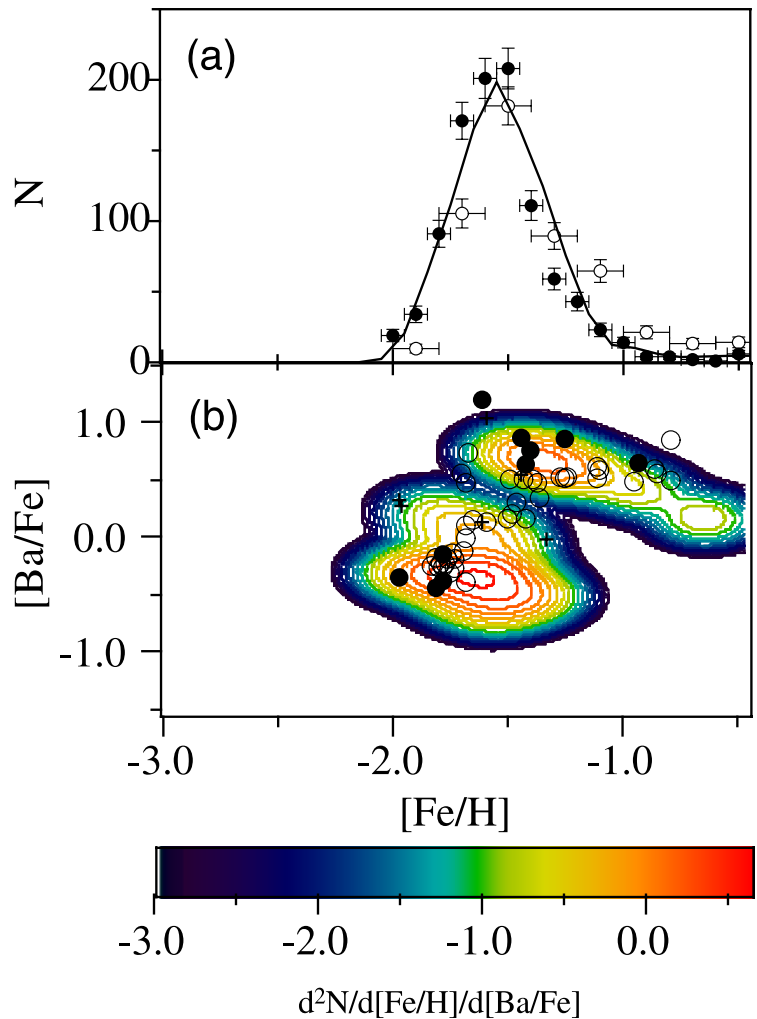}
\end{center}
\caption{($a$) Frequency distribution of $\omega$ Cen stars
against iron abundance, compared with observations ({\it filled
circles}; van Leeuwen et al.~2000, {\it open circles}; Norris et
al.~1996). The values of [Ca/H] for each star are converted to [Fe/H]
using the [Ca/Fe] ratios as a function of [Fe/H] shown in Figure
1. The error bars take into account Poisson noise. ($b$) Color-coded
frequency distribution of $\omega$ Cen stars in the [Ba/Fe]-[Fe/H]
plane convolved with a Gaussian with $\sigma=0.2$ dex for [Ba/Fe] and
$\sigma=0.1$ dex for [Fe/H]. The symbols represent the data taken from
Smith et al.~2000 ({\it filled circles}), Norris \& Da Costa 1995
({\it open circles}), and Smith, Cunta, \& Lambert 1995 ({\it
pluses}).}
\end{figure*}

\section{Modelling of Chemical Evolution in $\omega$ Cen}

The chemical evolution of $\omega$ Cen is treated using an
inhomogeneous chemical evolution model \citep{Tsujimoto2_99} that
accounts for the star formation history described in the previous
section. The model is based on a scenario in which the chemical
evolution proceeds through a repetition of a sequence of SN explosion,
shell formation, and star formation from the matter swept up by
individual SNRs. Heavy elements ejected from an SN are assumed to be
trapped and well-mixed within the SNR shell. A fraction of this heavy
element material will be incorporated into stars of the next
generation, while the material remaining in the gas will mix with the
ambient medium. The above process will repeat, increasing metallicity
until SNRs can no longer sweep up enough gas to form shells. No stars
will form from SNRs after this happens, and the process will
terminate. Details of this process are given in
\citet{Tsujimoto2_99}. It should be noted that $\omega$ Cen stars are
born from remnants of not only SNe II but also SNe Ia. The lifetime of
SN Ia progenitors is assumed to be 1--3 Gyr. This lifetime can be
determined from the observed break in [O/Fe] at [Fe/H]$\sim -1$ in the
solar neighborhood \citep{Yoshii_96} with the help of the
age-metallicity relation. SNe Ia starts ejecting Fe at [Fe/H]$\sim -1$
to decrease the [O/Fe] ratios.

The metallicity of metal-poor stars is determined by the mass swept up
by each SNR together with SN yields. This mass is controlled by the
velocity dispersion of gas (TS02). As a result, the observed ADF of
these stars constrains the velocity dispersion. It was found that the
velocity dispersion $\sigma_{\rm v}$=36 km s$^{-1}$ explains the
observed ADF in the metal-poor region. A smaller velocity dispersion
produces more metal-poor stars than observed, and vice
versa. Therefore a constant velocity dispersion of 36 km s$^{-1}$ is
assumed.

The evolution of Fe in $\omega$ Cen can then be calculated using this
information. The free parameters in the model are the mass fraction
$X_{\rm init}$ of stars initially formed in the first two star
formation phases, and the mass fraction $\epsilon$ of stars formed in
the dense shell swept up by each SNR. The mass fraction $X_{\rm init}$
is assumed to be $2.5\times10^{-4}$, as adopted for halo field stars
\citep{Tsujimoto2_00}. This value is assumed in both star formation
phases because the results are largely insensitive to this value as
long as $X_{\rm init}<1$\%. The value of $\epsilon$ is determined so
as to reproduce the observed ADF. The shape of the ADF for [Fe/H]$<-1$
requires $\epsilon$=$1.4\times 10^{-2}$ (first) and $1.5\times
10^{-2}$ (second). These combinations of $X_{\rm init}$ and $\epsilon$
give star formation durations of $\sim 250$ Myr (first) and $\sim 200$
Myr (second). The interval between these two star formation phases is
set at $\sim 300$ Myr to account for the predicted enrichment by
$s$-process elements in the meanwhile. A longer interval of $\sim 500$
Myr is allowed from this condition as long as SNe Ia do not contribute
to chemical enrichment in the second episode. The lower limit for the
fraction of gas lost due to SN explosion during the interval is
obtainable from the [Ba/Eu] ratios in stars with [Fe/H]$>-1.6$. This
condition gives a fraction of 70\%, which is used in the following
calculations. For the SN Ia-induced star formation,
$\epsilon$=$1.7\times 10^{-2}$ is adopted to reproduce the extended
metal-rich tail. As a result of calculations, the following star
formation history for $\omega$ Cen is proposed.

The proto-cluster cloud for $\omega$ Cen was initially enriched to
[Fe/H]=$-2.0$ with [Ba/Fe]=$-0.4$, corresponding to that of halo field
stars at the same [Fe/H]. From this enriched gas, star formation
proceeded for $\sim 250$ Myr followed by a quiescent period of $\sim
300$ Myr. The second star formation epoch continued for $\sim 200$
Myr, followed $\sim 250$ Myr later by an SN Ia event that initiated SN
Ia-induced star formation, which lasted for the next few Gyr. Figure
2a shows the ADF obtained for this model in comparison with the data
acquired by \citet{van Leeuwen_00} ({\it filled circles}) and
\citet{Norris2_96} ({\it open circles}). The result has been convolved
with a Gaussian with dispersion of $\sigma$=0.1 dex, which is
identical to the measurement error for [Fe/H]. The ADF for the
proposed model is in good agreement with the data.

The total fraction of gas converted into stars is $\sim$ 10\% of the
initial gas. It follows from the present mass of $\omega$ Cen
$\sim3\times 10^6$\ms \citep{Meylan_86} that the initial mass of the
proto-cluster cloud of $\omega$ Cen was $\sim 3\times10^7$\msp. From
this initial mass, it is possible to estimate the energy supplied by
SNe after the end of star formation in the first episode. The energy
from the preceding supernovae is lost from the system by
radiation. From the result that $\sim$ 7\% of the initial mass was
converted to stars in the first episode, the total energy added by
these last SNe amounts to $\sim 8\times10^{53}$ erg. The gravitational
binding energy of the gas is estimated by the virial theorem to be
$E_{\rm g}=GM^2/R=M\sigma_{\rm v}^2\sim8\times 10^{53}$ erg, which is
comparable to the thermal energy. Thus, part of the gas is expected to
be lost from the system. Here, as discussed above, the evolution of
[Ba/Eu] requires that 70\% of the gas was lost due to SNe II
explosion. An upper bound of the expelled gas fraction can be deduced
from the result that the number of stars produced in the second star
formation episode needs to be about a quarter of that produced in the
first episode to reproduce the observed ADF, which is in good accord
with the claim by \citet{Norris2_96}.  As a consequence, the fraction
of remaining gas must have been more than 25\%, that is, less than
75\% of the gas was lost. Thus the assumed value of 70\% satisfies
this criterion. From the remaining gas, second stage of enrichment
proceeded due to $\sim 2000$ SNe II in total, which is then followed
by third stage induced by $\sim 2500$ SNe Ia.

For the evolution of Ba in $\omega $ Cen, it is assumed that SNe II
with $M_{\rm ms}=20-25M_\odot$ are the dominant sites for $r$-process
nucleosynthesis \citep{Tsujimoto2_00, Tsujimoto_01}, and 2--4 \ms AGB
stars are the most likely sites for $s$-process nucleosynthesis
\citep{Travaglio_99}. Figure 2b is a color-coded frequency
distribution of stars in the [Ba/Fe]-[Fe/H] plane, normalised to unity
when integrated over the entire area. In order to make a direct
comparison with the data, the frequency distribution has been
convolved with a Gaussian with $\sigma$= 0.2 dex for [Ba/Fe] and
$\sigma$=0.1 dex for [Fe/H]. The result also covers the present
observational data points.

\section{$\omega$ Centauri as a Globular Cluster}

\begin{figure*}[htb]
\begin{center}
\includegraphics[width=15cm,angle=0]{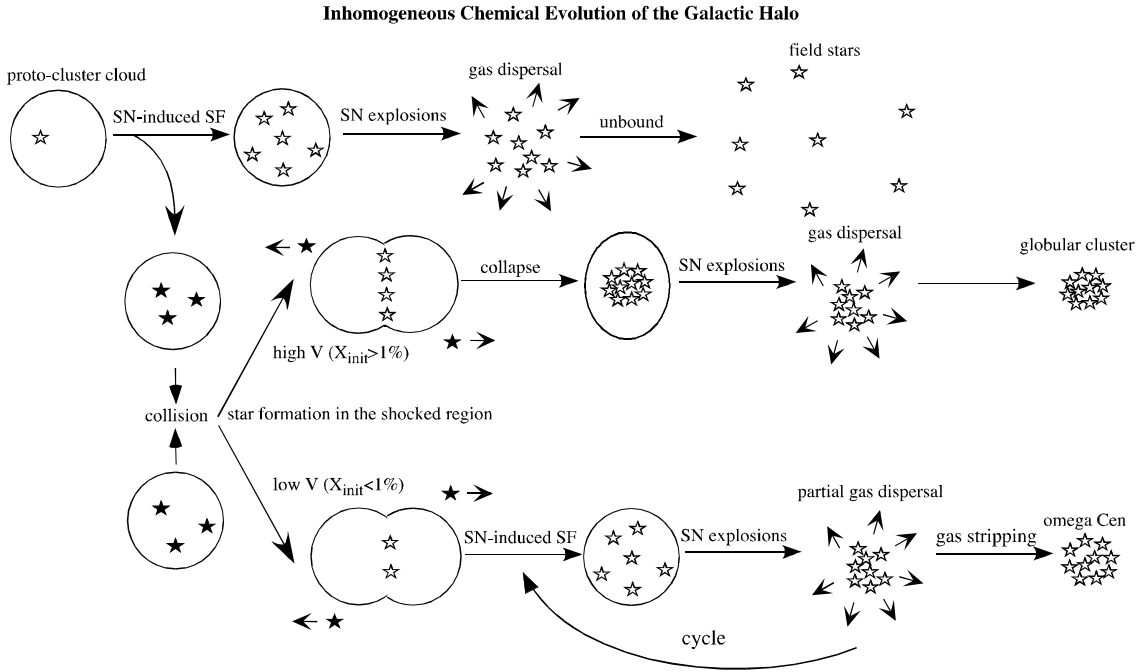}
\end{center}
\caption{Schematic picture of inhomogeneous chemical
evolution model for the Galactic halo.}
\end{figure*}

The quantitative features obtained for $\omega$ Cen in the previous
section can then be discussed in the context of GC formation. We
propose that GCs formed through collisions of the proto-cluster clouds
at the halfway of SN-induced star formation. The scheme of GC
formation discussed here is described in Figure 3. The key points of
this scenario are summarised in this section. Details including the
comparison of the ADF calculated using this model with observations
have been presented in \citet{Shigeyama_03}.

\subsection{Pre-history erased by cloud-cloud collision}

The absence of stars with [Fe/H]$<-2.5$ in GCs suggests that the
proto-cluster clouds already possessed a certain abundance of heavy
elements. A likely mechanism for the formation of such clusters is
cloud-cloud collision, in which each cloud undergoes star formation
\citep[and references therein]{Murray_93}. Cloud-cloud collisions can
leave shocked gas with some heavy elements but without stars or dark
matter due to the expected high relative velocity ($\sim$200 km
s$^{-1}$), at which the stars and dark matter in each cloud would
proceed without being affected by the gravity from these clouds.

\subsection{Star formation after merging}

After two clouds collide, strong shock waves are formed, which
compress the gas and leave a dense cloud with sufficient mass for star
formation if the collision is nearly head-on. From this compressed
gas, stars are assumed to form with metallicity inherited from the
gas. Hydrodynamical instabilities (e.g., the Richtmeyer-Meshkov
instability) occurring at the interface of two clouds induce turbulent
motions. The turbulent diffusion will induce chemical homogeneity in
these clouds \citep{Murray_90}.  In the SN-driven star formation
scenario, the star formation history is determined by the evolution of
SNRs, which is affected by the environment. A critical indicator of
chemical evolution is whether star formation proceeded for a number of
generations or ceased after only one or a few generations. This
condition is given by comparing the maximum size that an SNR can
attain and the size of the cloud. If the maximum size of an SNR
exceeds the size of the cloud, the SNR cannot assemble the gas
necessary to form stars of the next generation. The resultant GC would
then contain a single generation of stars with no dispersion in
[Fe/H]. If the SNR is smaller than the cloud, the cloud would contain
multiple generations of stars with different [Fe/H], resembling
$\omega$ Cen stars. An additional condition for the formation of
multiple generations of stars is that the mass fraction $X_{\rm init}$
of stars in a cloud at the beginning should be less than 1\%, as
discussed previously \citep{Tsujimoto2_99}. If $X_{\rm init}$ is
larger than 1\%, the total gas swept up by the first SNRs exceeds the
entire amount of available gas, and thus star formation stops at the
first or with a few generations.

For a quantitative and simple discussion, a head-on collision of two
identical clouds is considered here. Suppose that each cloud moving at
speed $v_1$ is spherical and uniform with mass $M=10^6\,\Msun$ and
radius $R=40$ pc. This mass corresponds to the Jeans mass soon after
recombination. The velocity dispersion $\sigma_{\rm v}$ of $\sim 10 $
km s$^{-1}$, accounting for the elemental abundance patterns of
extremely metal-poor field stars \citep{Shigeyama_98}, allows $R$ to
be estimated. The density $\rho_2$ of the strongly shocked region is
given by
\begin{equation}
\rho_2=\rho_1\left(1+\frac{2}{\sqrt{1+\frac{4P_1}{\rho_1v_1^
2}}-1}\right)\sim\rho_1\left(\frac{\rho_1 v_1^2}{P_1}\right),
\end{equation}
where $P_1=2\sigma_{\rm v}^2\rho_1/3$ is the pressure in the
pre-shocked gas, and $\rho_1$ denotes the initial mean density of the
cloud. Here, the shock is assumed to be isothermal because the cooling
time is several orders of magnitude shorter than the sound crossing
time. The expected isochoric cooling will prevent the merging clouds
from expanding after the shock waves reach the surface. As a result,
the collision leaves a cloud with a disk-like shape with a half-width
of $W=2R\times(\rho_1/\rho_2)=4GM/(3v_1^2)$. The shocked cloud may not
feed an SNR if the size of an SNR given by $R_{\rm SNR} \sim 0.9\,{\rm
pc}\left(v_1/(200\,{\rm km\,s}^{-1})\right)^{-2/3}$
\citep{Shigeyama_98} exceeds the width of the cloud. Thus, the cloud
with $R_{\rm SNR}>W$ will produce a GC with a single generation of
stars, while a cloud with $R_{\rm SNR}<W$ will be able to produce a GC
with multiple star generations. This criterion gives the maximum
velocity that can produce a GC with some dispersion in [Fe/H] for this
particular parameter set, as follows,
\begin{equation}
v_{1,\,\rm
max}\sim 33 \,{\rm km \ s}^{-1}\left(\sigma_{\rm v}/10\,{\rm
km\,s}^{-1}\right)^{37/28}\left(M/10^6\Msun\right)^{1/4}.
\end{equation}
\noindent
Since clouds move at $\sim 200$ km s$^{-1}$ in the gravitational
potential of the Milky Way, it is rare that two clouds would collide
with such low velocity. In fact, if all the clouds move at $200$ km
s$^{-1}$ and their distribution is isotropic in the velocity space,
then the fraction of collisions with $v_1<33$ km s$^{-1}$ is $\sim
0.027$. The width of the compressed cloud as small as $W\sim 0.15
\,{\rm pc}$ induces chemical homogeneity on a short timescale of $\sim
10^4$ yr.

On the other hand, the mass and velocity dispersion of the
proto-cluster cloud generating $\omega$ Cen have been estimated to be
$M\sim3\times10^7$\msp, $\sigma_{\rm v} \sim 36$ km s$^{-1}$ from the
chemical evolution model presented in section 3. These values together
with equation (2) suggest that a supernova does not break up the
shocked clouds if two proto-cluster clouds collide with relative
velocity ($2 v_1$) of less than $\sim 700$ km s$^{-1}$. Therefore,
$\omega$ Cen-like clusters are most likely to form from two colliding
massive clouds if $X_{\rm init}$ is less than 1\%. Such a small
$X_{\rm init}$ could be realized by a small relative velocity leading
to weak shock-wave compression. After a small mass fraction ($X_{\rm
init}<1$\%) of first-generation stars had formed in colliding
proto-cluster clouds, chemical enrichment due to SN-induced star
formation would then have proceeded in $\omega$ Cen. A small relative
velocity at the collision may lead to chemical inhomogeneity in the
compressed proto-cluster cloud of $\omega$ Cen, which might be
partially responsible for the observed metallicity dispersion.

\subsection{Final stage}

When SNe from the first generation stars break up the gas, it is
likely that the available energy from SNe exceeds the gravitational
binding energy of the cloud. Since the dynamical timescale of the
cloud discussed above ($\sim$4 Myr) is shorter than that for the
evolution of SN II progenitor stars, the stellar component with velocity
dispersion much smaller than that of the gas will shrink to form a
bound system before SNe expel the remaining gas. In fact, the virial
theorem implies that the half mass radius $R_{\rm h}$ of the
newly formed GC is $R_{\rm h}\sim GM_{\rm GC}/(4\sigma_{\rm v}^2) \sim
2.2$ pc. This value is significantly smaller than the initial radius
and is in accordance with the observation.

When the star formation repeats for generations as in the
proto-cluster cloud for $\omega$ Cen, the available energy from SNe
must be counted to specify the time of gas removal, as was done for
TS02. This energy is found to be insufficient for total gas removal as
already discussed in the previous section. Subsequent passage through
the Galactic gaseous disk is an alternative process that may have
stripped away the remaining gas. This ram pressure stripping was not
efficient during the star formation phase in $\omega$ Cen because the
gas density in the forming Galactic disk during the first few Gyrs is
quite small, which is attributed to a slow formation of the Galactic
disk inferred from the chemical evolution in the solar neighborhood
\citep{Yoshii2_96}. The gas density $\rho_{\rm disk}$ of the forming
disk at the age of 2 Gyr for instance is estimated to be $\sim$ 1/4 of
the present density. From the present $\rho_{\rm disk}$ of $\sim 0.04$
\ms pc$^{-3}$ \citep{Cox_00} and the gas density $\rho_\omega$ of the
proto-cluser cloud for $\omega$ Cen during the second and third
episodes of $\sim 2$ \ms pc$^{-3}$, the timescale $t_{\rm rp}$ for the
ram pressure stripping of $\omega$ Cen, defined as $t_{\rm rp}\approx
R/v(2\rho_\omega/\rho_{\rm disk})^{1/2}$ \citep{Sarazin_88}, is $\sim
10^7$ yr, where $R$ and $v$ are the radius ($\sim$100 pc) and the
velocity ($\sim$200 km s$^{-1}$) of the proto-cluster  cloud for
$\omega$ Cen. On the other hand, the timescale $t_{\rm pass}$ for the
passage through the forming disk is estimated to be $\sim 3\times10^6$
yr, assuming the present disk scale height of 350 pc. Therefore the
ram pressure is not strong enough to strip the gas but is expected to
induce a bulk motion of the gas. The gradual increase in the gas
density in the forming disk has made $t_{\rm rp}$ comparable to
$t_{\rm pass}$ in the next few Gyrs. As a result, the remaning gas
after the third episode of star formation in $\omega$ Cen is likely to
be stripped by the interaction with the forming Galactic disk. In this
connection, we will refer to a possibility of the gas stripping by
tidal interaction during the star formation phase in $\omega$
Cen. Even if $\omega$ Cen originally had the orbit with a pericenter
of 1 kpc \citep{Dinescu_99}, the tidal radius was $\sim$100 pc,
comparable to the radius of the proto-cluster cloud for $\omega$
Cen. Furthermore, the orbit might have decayed to the present orbit as
discussed by \citet{Zhao_02}.  The pericenter during the first few
Gyrs could be $\sim$7 kpc. The corresponding tidal radius of $\omega$
Cen becomes several hundreds pc. Therefore the tidal stripping is not
significant for the evolution of $\omega$ Cen.

\section{CONCLUSIONS}

A model that successfully reproduces the observed chemical properties
in $\omega$ Cen was constructed. The model suggests that the GC
$\omega$ Cen originated from a cloud-cloud collision that triggered
the birth of first-generation stars. Each of the clouds had a mass of
$\sim1.5\times 10^7\,\Msun$ and size estimated from the velocity
dispersion of $\sim100$ pc. The relative velocity of the collision
must have been low enough to limit the mass fraction of the
first-generation stars to less than 1\%. If this wasn't the case, no
dispersion in [Fe/H] would be seen in the member stars. The first star
formation lasted for $\sim$ 250 Myr, enriching the gas with heavy
elements to [Fe/H]$\sim -1.6$. Supernovae at the end of this epoch
were unable to assemble the remaining gas to form stars or expel all
the gas. Instead, $\sim 30$\% of the gas remained bound and star
formation recommenced $\sim 300$ Myr after the end of the first star
formation epoch. Between the two epochs, AGB stars began to
contaminate the remaining gas with $s$-process elements up to
[Ba/Fe]$\sim +1$. About 200 Myr later, star formation ceased.
Supernovae at the end of this period also failed to expel the
gas. Eventually, SNe Ia initiated SN-induced star formation,
exclusively supplying Fe to the gas.

This study of $\omega$ Cen revealed a new, different picture of the GC
formation scenario. In this scenario, first-generation stars are born
through a cloud-cloud collision, and the formation of later-generation
stars occurs only when the dense shell of SNRs originating from the
first-generation SNe is formed. Such a situation is realized only when
the relative velocity of the colliding massive clouds is
low. Otherwise, only a single generation of stars will be created in
the GCs, resulting in no spread in [Fe/H]. In the sense that the
massive proto-cluster clouds must be rare, as would be low-velocity
collisions, $\omega$ Cen should indeed be part of a minor population.

\acknowledgements

We are grateful to the anonymous referee for making useful comments.
This work has been partly supported by COE research (07CE2002) and a
Grant-in-Aid for Scientific Research (11640229, 12640242) of the
Ministry of Education, Culture, Sports, Science, and Technology in
Japan.


\begin{thebibliography}{}
\bibitem[Cox(2000)]{Cox_00} 
Cox, A. N. 2000, Allen's Astrophysical Quantities Fourth Edition.  
Springer-Verlag New York
\bibitem[Dinescu, Girard, \& van Altena(1999)]{Dinescu_99}
Dinescu, D. I., Girard, T. M., \& van Altena, W. F. 1999, ApJ, 117, 1792 
\bibitem[Ferraro, Bellazzini, \& Pancino(2002)]{Ferraro_02}
Ferraro, F. R., Bellazzini, M., \& Pancino, E. 2002, ApJ, 573, L95
\bibitem[Freeman(1993)]{Freeman_93}
Freeman, K. C. 1993, in ASP Conf. Ser. 48, The Globular Cluster-Galaxy 
Connection, ed. G. H. Smith \& J. P. Brodie (San Francisco: ASP), 608
\bibitem[Gnedin et al.(2002)]{Gnedin_02}
Gneden, O. Y., Zhao, H., Pringle, J. E., Fall, S. M., Livio, M., \&
Meylan, G. 2002, ApJ, 568, L23
\bibitem[Hilker \& Richter(2000)]{Hilker_00}
Hilker, M. \& Richter, T. 2000, A\&A, 362, 895
\bibitem[Ikuta \& Arimoto(2000)]{Ikuta_00}
Ikuta, C., \& Arimoto, N. 2000, A\&A, 358, 535
\bibitem[Kraft(1979)]{Kraft_79}
Kraft, R. P. 1979, ARA\&A, 17, 309
\bibitem[Majewski et al.(2000)]{Majewski_00}
Majewski, S. R., Patterson, R. J., Dinescu, D. I., Johnson, W. Y.,
Ostheimer, J. C., Kunkel, W. E., \& Palma, C. 2000, in Proc. Liege
Int. Astrophys. Colloq., The Galactic Halo: From Globular Clusters to
Field Stars, ed. A. Noels et al., 619
\bibitem[Meylan \& Mayor(1986)]{Meylan_86}
Meylan, G., \& Maylor, M. 1986, A\&A, 166, 122
\bibitem[Murray \& Lin(1990)]{Murray_90}
Murray, S. D., \& Lin, D. N. C. 1990, \apj, 357, 105
\bibitem[Murray \& Lin(1993)]{Murray_93}
Murray, S. D., \& Lin, D. N. C. 1993, in The Globular Cluster-Galaxy
Connection, ASP Conference Series, Vol. 48, eds. G. H. Smith, \& J. P.
Brodie, 738
\bibitem[Norris \& Da Costa(1995)]{Norris_95}
Norris, J. E., \& Da Costa, G. S. 1995, ApJ, 447, 680
\bibitem[Norris, Freeman, \& Mighell(1996)]{Norris_96}
Norris, J. E., Freeman, K. C., \& Mighell, K. J. 1996, ApJ, 462, 241
\bibitem[Pancino et al.(2000)]{Pancino_00}
Pancino, E., Ferraro, F. R., Bellazzini, M., Piotto, G., \& Zoccali, M. 
2000, ApJ, 534, L83
\bibitem[Pancino et al.(2002)]{Pancino_02}
Pancino, E., Pasquini, L., Hill, V., Ferraro, F. R., \& Bellazzini, M.
2002,
ApJ, 568, L101
\bibitem[Sarazin(1988)]{Sarazin_88}
Sarazin, C. L. 1988, X-ray emission from clusters of galaxies 
(Cambridge: Cambridge Univ. Press)
\bibitem[Shigeyama \& Tsujimoto(1998)]{Shigeyama_98}
Shigeyama, T., \& Tsujimoto, T. 1998, \apj, 507, L135
\bibitem[Shigeyama \& Tsujimoto(2003)]{Shigeyama_03}
Shigeyama, T., \& Tsujimoto, T. 2003, in the APS conf. ser., ed.
S. Ikeuchi, J. Hearnshaw, \& T. Hanawa, in press
\bibitem[Smith(1987)]{Smith_87}
Smith, G. H. 1987, PASP, 99, 67
\bibitem[Smith, Cunta, \& Lambert(1995)]{Smith_95}
Smith, V. V., Cunta, K., \& Lambert, D. L. 1995, AJ, 110, 2827
\bibitem[Smith et al.(2000)]{Smith_00}
Smith, V. V., Suntzeff, N. B., Cunha, K., Gallino, R., Busso, M.,
Lambert, D. L., \& Straniero, O. 2000, AJ, 119, 1239
\bibitem[Suntzeff \& Kraft(1996)]{Suntzeff_96}
Suntzeff, N. B., \& Kraft, R. P. 1996, AJ, 111, 1913
\bibitem[Travaglio et al.(1999)]{Travaglio_99}
Travaglio, C., Galli, D., Gallino, R., Busso, M., Ferrini, F., \&
Straniero, O. 1999, ApJ, 521, 691
\bibitem[Tsujimoto \& Shigeyama(1998)]{Tsujimoto_98}
Tsujimoto, T., \& Shigeyama, T. 1998, \apj, 508, L151
\bibitem[Tsujimoto, Shigeyama, \& Yoshii(1999)]{Tsujimoto_99}
Tsujimoto, T., Shigeyama, T., \& Yoshii, Y. 1999, \apj, 519, L63
\bibitem[Tsujimoto, Shigeyama, \& Yoshii(2000)]{Tsujimoto_00}
Tsujimoto, T., Shigeyama, T., \& Yoshii, Y. 2000, \apj, 531, L3
\bibitem[Tsujimoto \& Shigeyama(2001)]{Tsujimoto_01}
Tsujimoto, T., \& Shigeyama, T. 2001, \apj, 561, L97
\bibitem[Tsujimoto \& Shigeyama(2002)]{Tsujimoto_02}
Tsujimoto, T., \& Shigeyama, T. 2002, \apj, 571, L93 (TS02)
\bibitem[van Leeuwen et al.(2000)]{van Leeuwen_00}
van Leeuwen, F., Le Poole, R. S., Reijns, R. A., Freeman, K. C.,
de Zeeuw, P. T. 2000, A\&A, 360, 472
\bibitem[Yoshii, Tsujimoto, \& Nomoto(1996)]{Yoshii_96}
Yoshii, Y., Tsujimoto, T., \& Nomoto, K. 1996, ApJ, 462, 266
\bibitem[Zhao(2002)]{Zhao_02} 
Zhao, H. S. 2002, in ASP Conf. Ser., Omega Centauri, A Unique Window into 
Astrophysics, ed. F. Van Leeuwen, G. Piotto, \& J. Huges (San Francisco: 
ASP), 391
\bibitem[Ferraro et al.(2002)]{Ferraro2_02}
\bibitem[Norris et al.(1996)]{Norris2_96}
\bibitem[Tsujimoto et al.(1999)]{Tsujimoto2_99}
\bibitem[Tsujimoto et al.(2000)]{Tsujimoto2_00}
\bibitem[Yoshii et al.(1996)]{Yoshii2_96}
\end{thebibliography}
\end{document}